\documentclass{modified_iaus}
\usepackage{graphicx,natbib}

\renewcommand{\vec}{\boldsymbol}

\title[Wreath-building dynamos] 
{Global-scale wreath-building dynamos in stellar convection zones}

\author[Brown, Browning, Brun, Miesch \& Toomre]   
{Benjamin P.\ Brown$^{1,2}$, 
 Matthew K.\ Browning$^3$, 
 Allan Sacha Brun$^4$, 
 Mark S.\ Miesch$^5$ \and
 Juri Toomre$^6$}
 
\affiliation{
  $^1$Dept.\ Astronomy, University of Wisconsin, Madison, WI 53706-1582\\
  email: {\tt bpbrown@astro.wisc.edu} \\[\affilskip]
  $^2$Center for Magnetic Self Organization in Laboratory and
  Astrophysical Plasmas, University of Wisconsin, Madison, WI 537066-1582 \\[\affilskip]
  $^3$Canadian Institute for Theoretical Astrophysics, University of Toronto, Toronto, ON M5S3H8 Canada \\[\affilskip]
  $^4$DSM/IRFU/SAp, CEA-Saclay and UMR AIM, CEA-CNRS-Universit\'e Paris 7, 91191 Gif-sur-Yvette, France \\[\affilskip]
  $^5$High Altitude Observatory, NCAR, Boulder, CO 80307-3000  \\[\affilskip]
  $^6$JILA and Dept.\ Astrophysical \& Planetary Sciences, University of Colorado, Boulder, CO 80309-0440}

\pubyear{2010}
\volume{271}  
\pagerange{1--8}
\jname{to be published in ``Astrophysical Dynamics: from Stars to Galaxies''}
\editors{ed.\ A.S.\ Brun, N.H.\ Brummell \& M.S.\ Miesch}
\begin{document}

\maketitle

\begin{abstract}
  When stars like our Sun are young they rotate rapidly and are very
  magnetically active.  We explore dynamo action in rapidly rotating
  suns with the 3-D MHD anelastic spherical harmonic (ASH) code.  The
  magnetic fields built in these dynamos are organized on
  global-scales into wreath-like structures that span the convection
  zone.  Wreath-building dynamos can undergo quasi-cyclic reversals of
  polarity and such behavior is common in the parameter space we have
  been able to explore.   These dynamos do not appear to require
  tachoclines to achieve their spatial or temporal organization.
  Wreath-building dynamos are present to some degree at all rotation
  rates, but are most evident in the more rapidly rotating simulations.
\end{abstract}

\firstsection 
\section{Introduction}
When stars like the Sun are young, they rotate quite rapidly.
Observations of these young Suns indicate that they have strong
surface magnetic activity and can undergo global-scale polarity
reversals similar to the 22-year solar cycle.  The magnetic fields
observed at the surface of these stars are thought to originate in
stellar dynamos driven in their convective envelopes.  There, 
plasma motions couple with rotation to generate global-scale magnetic
fields.  Though correlations between the rotation rate of stars and
their magnetic activity are observed
\citep[e.g.,][]{Pizzolato_et_al_2003} it is at present unclear how
the stellar dynamo process depends in detail on rotation.

Motivated by this rich observational landscape, we have explored the
effects of more rapid rotation on 3-D convection and dynamo action in
simulations of stellar convection zones.  These simulations have been
conducted using the anelastic spherical harmonic (ASH) code, a tool
developed by a team of postdocs and graduate students working with 
Juri Toomre to study global-scale magnetohydrodynamic convection and
dynamo action in stellar convection zones \citep[e.g.,][ and contribution by
Miesch in these proceedings]{Clune_et_al_1999, Miesch_et_al_2000,Brun_et_al_2004}.

We began our explorations of convection in rapidly rotating suns by
exploring hydrodynamic simulations at a variety of rotation rates
\citep{Brown_et_al_2008}. These simulations capture the convection
zone only, spanning from $0.72\:R_\odot$ to $0.97\:R_\odot$, and take
solar values for luminosity and stratification but the rotation rate
is more rapid. The total density contrast across such shells is about 25.
In those simulations we found that the differential rotation generally
becomes stronger as the rotation rate increases, while the meridional
circulations appear to become weaker and multi-celled in both radius
and latitude.   

In this paper we review the dynamos we have found in our simulations
of more rapidly rotating solar-type stars.  These wreath-building
dynamos form surprisingly organized structures in their convection
zones (\S\ref{sec:case D3}) and some even undergo quasi-cyclic
magnetic reversals (\S\ref{sec:case D5}).  Wreath-building dynamos
appear throughout the parameter space we have surveyed to date
(\S\ref{sec:parameter space}).  We close by reflecting on the
challenges that lie ahead (\S\ref{sec:challenges}).

\section{A Dynamo with Magnetic Wreaths}
\label{sec:case D3}

Our first simulation discussed here, case D3, is of a star rotating three times
faster than the Sun \citep{Brown_et_al_2010a}.  
Vigorous convection in this simulation drives a strong differential rotation and
achieves sustained dynamo action at relatively low magnetic Prandtl
number; here Pm=$\nu/\eta$ is 0.5, where $\nu$ is the viscosity and
$\eta$ is the magnetic diffusivity.

\begin{figure}[t]
\begin{center}
 \includegraphics[width=\linewidth]{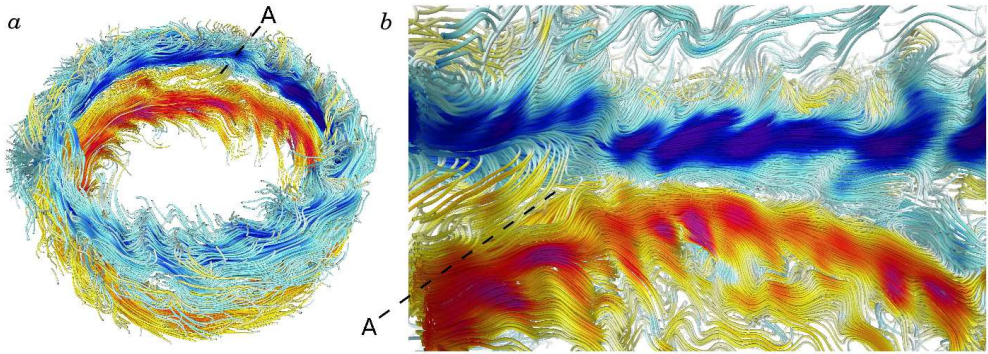}
 \vspace*{0.25 cm}
 \caption{Magnetic wreaths in case D3.  $(a)$ Full volume rendering of
   magnetic wreaths, showing entire simulation.  Lines trace the vector
   magnetic field with color denoting amplitude and polarity of the
   longitudinal magnetic field  $B_\phi$ (red or light tones,
   positive; blue or darker tones, negative).  Rather than being
   simple flux surfaces, magnetic fields thread in and out of each
   wreath, connecting the wreaths across the equator and linking them
   to the polar regions. $(b)$ Zoom in view of region A showing
   cross-equatorial connectivity.} 
   \label{fig:case D3}
\end{center}
\end{figure}

The magnetic fields created in this dynamo are organized on
global-scales into banded wreath-like structures. These are shown for
case~D3 in Figure~\ref{fig:case D3}$a$.  Two such wreaths are visible in
the equatorial region, spanning the depth of the convection zone and 
latitudes from roughly $\pm30^\circ$.  The dominant component of the
magnetism is the longitudinal field $B_\phi$, and the two wreaths have
opposite polarities.  Here the wreath in the northern hemisphere has
negative polarity $B_\phi$ while the wreath in the southern hemisphere
is positive in sense.
The wreaths are not isolated flux structures; instead, magnetic fields
meander in and out of each wreath, connecting them across the
equator and to higher latitudes (Fig.~\ref{fig:case D3}$b$).  The lack
of visible magnetism in the polar regions reflects the relatively low
magnetic Reynolds number associated with the convection 
(average fluctuating Rm' is roughly 50 at mid-convection zone).

It has been a great surprise that such structures can exist in the
convection zone of this simulation.  Generally, it has been expected
that convection should shred such structures or pump them downwards
into a stable tachocline at the base of the convection zone.  Here the
entire domain is convectively unstable, and no such tachocline is
present.  The wreaths persist for long intervals in time, with the
mean (longitudinally averaged) magnetic fields showing relatively
little variation in time.  The convection leaves its imprint on the
wreaths, with the strongest downflows dragging the field towards the
bottom of the convection zone.  This is visible in the distinct
waviness apparent in Figure~\ref{fig:case D3}.  On the poleward edges
the wreaths are wound up into the vortical convection there, and this
appears to play an important role in regenerating the poloidal field.

The magnetic wreaths are built by both the global-scale differential rotation
and by the emf arising from correlations in the turbulent convection.
Generally, the mean longitudinal magnetic field $\langle B_\phi
\rangle$ in the wreaths is generated by the $\Omega$-effect: the
stretching of mean poloidal field by the shear of differential
rotation into mean toroidal field.  Production of $\langle B_\phi
\rangle$ by the differential rotation is balanced by turbulent shear
and advection, and by ohmic diffusion on the largest scales.

The mean poloidal field is generated by the turbulent emf 
$E_\mathrm{FI} = \langle \vec{u'} \times \vec{B'} \rangle$, where the
fluctuating velocity is 
$\vec{u'}=\vec{u}-\langle \vec{u}\rangle$ and the fluctuating magnetic
fields are $\vec{B'}=\vec{B}-\langle \vec{B}\rangle$.  In these
simulations, $E_\mathrm{FI}$ is generally strongest at the poleward
edge of the wreaths, centered at approximately $\pm 20^\circ$ latitude,
whereas the $\Omega$-effect and $\langle B_\phi \rangle$ peak at
roughly $\pm 15^\circ$ latitude.  This spatial offset 
between $E_\mathrm{FI}$ and $\langle B_\phi \rangle$ means that the
turbulent emf is not generally well represented by a simple
$\alpha$-effect description, e.g.,
\begin{equation}
  E_\mathrm{FI} = \langle \vec{u'} \times \vec{B'} \rangle|_\phi \neq \alpha
  \langle B_\phi \rangle
\end{equation}
when $\alpha$ is a scalar quantity.  This is true even when $\alpha$
is estimated from the kinetic and magnetic helicities
present in the simulation.  More sophisticated mean-field models may
do much better at matching the observed emf $E_\mathrm{FI}$, and other
terms in the mean-field expansion may play a significant role; in
particular, the gradient of $\langle B_\phi \rangle$ is large on the
poleward edges of the wreaths where $E_\mathrm{FI}$ is significant.

\section{A Cyclic Dynamo in a Stellar Convection Zone}
\label{sec:case D5}

We turn now to a more rapidly rotating dynamo simulation, case~D5,
rotating five times faster than the Sun \citep{Brown_et_al_2010b}.  As
in case~D3, strong global-scale magnetic wreaths are built in the
convection zone.  Now however, the wreaths begin to show significant
time-variation and undergo quasi-regular polarity reversals.

\begin{figure}[t]
\begin{center}
 \includegraphics[width=\linewidth]{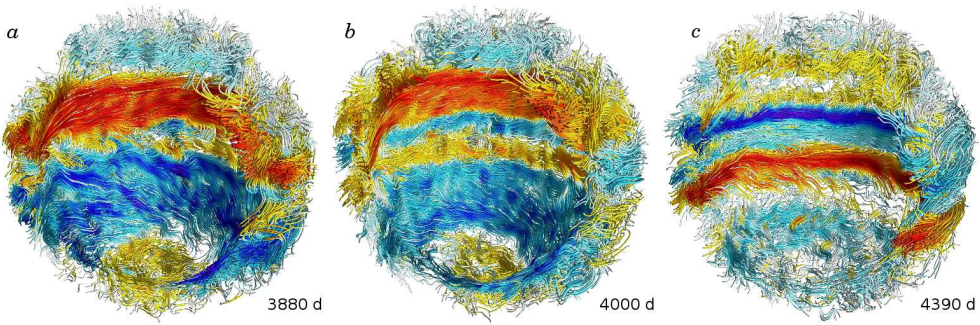}
 \vspace*{0.25 cm}
 \caption{Global-scale magnetic reversal in case D5.
   $(a)$ Magnetic wreaths shown in field line tracing shortly before a
   polarity reversal, with positive polarity wreath above equator and
   negative polarity below. Volume shown spans slightly more than a full
   hemisphere with both polar caps visible where relic wreaths from
   the previous cycle remain visible. $(b)$~During a reversal,
   new wreaths with opposite polarity 
   form at the equator while old wreaths migrate toward
   the poles.  $(c)$ When the reversal completes, the polarity of the
   wreaths have flipped, with negative polarity wreath above the
   equator and positive below.  Cancellation occurs in the polar
   regions and the old wreaths slowly dissipate.
   Times of snapshots are labeled.}
   \label{fig:case D5}
\end{center}
\end{figure}

One such reversal is illustrated in Figure~\ref{fig:case D5}.
Before the reversal (Fig.~\ref{fig:case D5}$a$), the wreaths look
similar to those found in case~D3, though here magnetism permeates 
the entire convection zone, including the polar regions where relic
wreaths from the previous reversal are visible.  The equatorial region
shows significantly more connectivity and large fluctuations of
$B_\phi$, with small knots of alternating polarity visible throughout.
This cross-equatorial connectivity appears to play an important role
in the reversal process.  The magnetic fields built in case~D5 attain
somewhat larger amplitudes than those realized in case~D3: here at
mid-convection zone $B_\phi$ can reach $\pm 40\:$kG,  
while in case~D3 the peak amplitudes were closer to $\pm 26\:$kG.

During the reversal (Fig.~\ref{fig:case D5}$b$), new wreaths of
opposite polarity form near the equator while the old wreaths
propagate towards the poles.  This poleward propagation appears to
be a combination of a nonlinear dynamo wave, arising from systematic
spatial offsets between the generation terms for mean poloidal and
toroidal magnetic field, and possibly a poleward-slip instability
arising from magnetic stresses within the wreaths.
After a reversal (Fig.~\ref{fig:case D5}$c$) the new magnetic wreaths
grow in strength and dominate the equatorial region.  In the polar
regions the wreaths from the previous cycle begin to dissipate,
reconnecting with the pre-existing flux there and being shredded by
the turbulent convection there.

Generally, the dynamo processes identified in case~D3 serve to build
the wreaths of case~D5.  As in case~D3, before a reversal the
turbulent emf $E_\mathrm{FI}$ that contributes to the mean poloidal
field is generally largest on the poleward edge of the wreaths at
latitudes above $\pm 20^\circ$, while the $\Omega$-effect and 
$\langle B_\phi \rangle$ peak at roughly latitude $\pm 15^\circ$.
During a reversal both $E_\mathrm{FI}$ and the production of $\langle
B_\phi \rangle$ associated with the $\Omega$-effect surf on the
poleward edge of the wreaths as those structures move poleward.  This
systematic phase shift appears to contribute to that propagation.

\section{Wreath-building Dynamos}
\label{sec:parameter space}

The two simulations we have explored here, cases D3 and D5, are part
of a much larger family of simulations that we have conducted
exploring convection and dynamo action in younger suns.  The properties
of this broad family are summarized in Figure~\ref{fig:parameter
  space}$a$.  Indicated here are 26 simulations at rotation rates ranging
from $0.5\:\Omega_\odot$ to $15\:\Omega_\odot$.  At individual rotation
rates (e.g., $3\:\Omega_\odot$), further simulations explore the
effects of lower magnetic diffusivity $\eta$ and hence higher magnetic
Reynolds numbers.  Some of these follow a path where the magnetic
Prandtl number Pm is fixed at 0.5 (triangles) while others sample up
to Pm=4 (diamonds).  The most turbulent simulations have fluctuating
magnetic Reynolds numbers of about 500 at mid-convection zone.
Wreath-building dynamos are achieved in most simulations (17), though
a smaller number do not successfully regenerate their mean poloidal
fields (9, indicated with crosses).  Very approximate regimes of
dynamo behavior are indicated, based on the time variations shown by
the different classes of dynamos.

\begin{figure}[t]
\begin{center}
 \includegraphics[width=\linewidth]{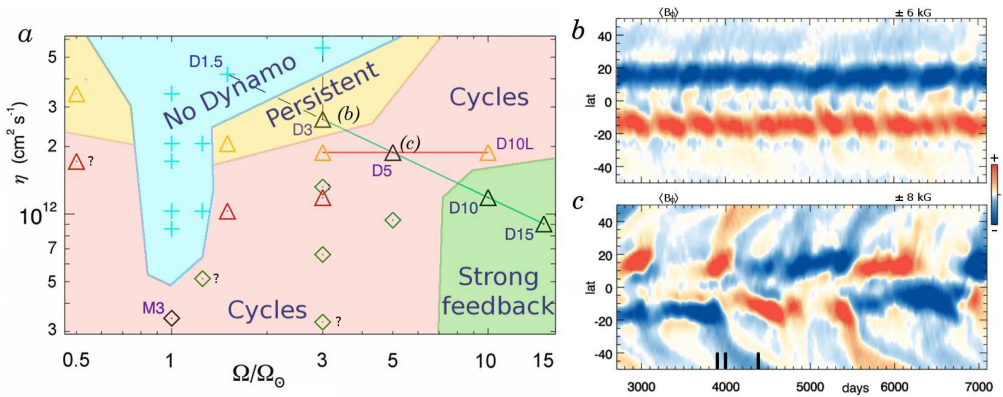}
\vspace*{0.25 cm}
 \caption{Wreath-building dynamos.  $(a)$ Parameter space showing
   variety of wreath-building dynamos currently explored.  Magnetic
   diffusivity $\eta$ and rotation rate $\Omega$ are shown for dynamo
   simulations at rotation rates ranging sampling
   $0.5$--$15\Omega_\odot$, with very approximate dynamo regimes shown.
   In some regions, magnetic Reynolds numbers are too low to sustain
   dynamo action, while in other regions persistent magnetic wreaths
   form which do not show evidence for cycles.  At higher magnetic
   Reynolds numbers (occurring here at low $\eta$ or high $\Omega$), 
   wreaths typically undergo quasi-cyclic
   reversals.  At the highest rotation rates the Lorentz force can
   substantially modify the differential rotation, but dynamo action
   is still achieved.  Cases with question marks show significant
   time-variation but have not been computed for long enough to definitively
   establish cyclic behavior.  $(b)$~Time-latitude plot of mean (axisymmetric)
   $B_\phi$ at mid-convection zone in persistent case D3
   \citep{Brown_et_al_2010a}.   
   $(c)$~Cyclic case D5 shown for same span of time
   \citep{Brown_et_al_2010b}.  Three reversals are visible here,
   occuring on roughly 1500 day periods.  Times of snapshots shown in
   Figure~\ref{fig:case D5} are indicated.
}
   \label{fig:parameter space}
\end{center}
\end{figure}

Near the onset of wreath-building dynamo action we generally find
little time variation in the axisymmetric magnetic fields associated
with the wreaths.  This is illustrated for case~D3 in
Figure~\ref{fig:parameter space}$b$, showing the mean longitudinal 
$\langle B_\phi \rangle$ at mid-convection zone over an interval of
about 4000 days.  Though small variations are visible on a roughly 500
day timescale, the two wreaths retain their polarities for the entire
time simulated (more than 20,000 days), which is significantly longer
than the convective overturn time (roughly 10--30 days), the rotation
period (9.3 days), or the ohmic diffusion time (about 1300 days at
mid-convection zone).  We refer to the dynamos in this regime as
persistent wreath-builders. 

Generally, we find that wreath-building dynamos begin to show large
time dependence as the magnetic diffusivity $\eta$ decreases and as
the rotation rate $\Omega$ increases.  In many cases this leads to
quasi-regular global-scale reversals of magnetic polarity, as
discussed for case~D5 in \S\ref{sec:case D5}.
We illustrate three of these cycles occuring in case~D5
in Figure~\ref{fig:parameter space}$b$.  In this simulation, reversals
occur with a roughly 1500 day timescale, though during some
intervals the dynamo can fall into other states.  For reference, the
ohmic diffusion time in this simulation is about 1800 days, while the
rotation period is 5.6 days.  Similar cycles occur in other
simulations, but the period of reversals varies with both $\eta$ and
$\Omega$.  Cycles appear to become shorter as $\eta$ decreases,
opposite to what might be expected if the ohmic time determined the
cycle period.  The dependence of cycle period on $\Omega$ is less
certain.  At present, determining why cycles are realized in many of these
dynamos remains difficult.  The phenomena appears to be at least
partially linked to the magnetic Reynolds number of the 
differential rotation and possibly to that of the fluctuating
convection. 

In these wreath-building dynamos the major reservoir of kinetic energy
that feeds the generation of magnetism is the axisymmetric
differential rotation, and this global-scale shear is strongly reduced
in the dynamo simulations. 
Individual convective structures are largely unaffected by the
magnetic wreaths except when the fields reach very large amplitudes;
in case D5 this occurs when $B_\phi$ exceeds values of roughly 35~kG
at mid-convection zone.
At the highest rotation rates the Lorentz force of the axisymmetric
magnetic fields becomes strong enough to substantially modify the
differential rotation, largely wiping out the latitudinal and radial
shear (e.g., cases D10 and D15 in Fig.~\ref{fig:parameter space}$a$).
In these cases, wreath-like structures can still form though they
typically have more complex structure and are less axisymmetric.

\begin{figure}[t]
\begin{center}
 \includegraphics[width=\linewidth]{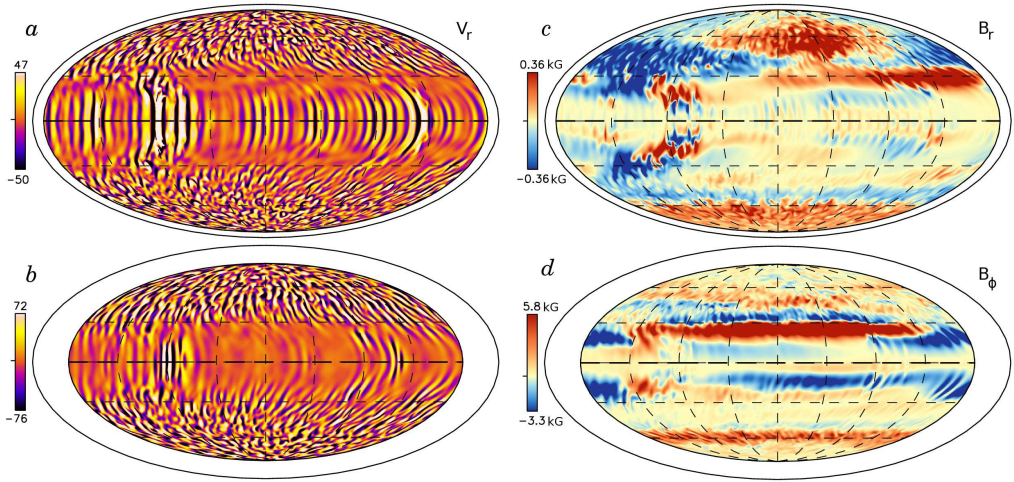}
\vspace*{0.25 cm}
 \caption{Nests of convection in dynamo case D10L.  
          Convective patterns are shown by snapshots in Mollweide
          projection of the radial velocities $v_r$ $(a)$~near the
          surface ($0.95\thinspace R_\odot$) and 
          $(b)$~at mid-convection zone
          ($0.85\thinspace R_\odot$).  Two active nests of
          convection are clearly visible.  $(c)$~Near the surface,
          the radial magnetic field $B_r$ is concentrated in the
          stronger nest.  This patch of radial field propagates with
          the nest.  $(d)$~At mid-convection zone the active
          nests leave their imprint on the magnetic wreaths, shown here by
          longitudinal field~$B_\phi$.
            \label{fig:case D10L}
}
\end{center}
\end{figure}

Some of these rapidly rotating dynamos have held further surprises.
One such simulation, case D10L rotating ten times faster than the Sun,
is shown in Figure~\ref{fig:case D10L}.  Magnetic wreaths form in this
dynamo, but so do strong localized nests of convection.  
Two such nests are visible near the equatorial region in the convective
patterns of radial velocity (Fig.~\ref{fig:case D10L}$a,b$).  At certain
longitudes the convection is significantly stronger, and these nests
span the convection zone.  The nests leave their imprint on the
magnetism, concentrating the radial magnetic field $B_r$ into patches
near the surface (Fig.~\ref{fig:case D10L}$c$) and interacting
strongly with the toroidal field $B_\phi$ at depth (Fig.~\ref{fig:case D10L}$d$). 
These active nests appear to be distinct dynamical structures, and
they propagate at a different rate than either individual convective
cells or the local differential rotation.  If such nests
persist in stellar convection zones, they are likely to strongly affect
inferences of the global-scale magnetic field. We have found active
nests in rapidly rotating hydrodynamic simulations 
\citep{Brown_et_al_2008}, but to our knowledge this is the first time
such structures have been found in a fully saturated dynamo state.

We have found wreath-like structures in simulations rotating at the
solar rotation rate as well (one example appears in the contribution by
Miesch, these proceedings).  Achieving such structures in solar simulations
is challenging, as the angular velocity contrast in the Sun is smaller
than that realized in the rapidly rotating dynamos.  Generally the
solar simulations require low values of $\eta$ to build wreaths, which
in turn calls for high resolutions and that exacts a large computational cost.

The magnetic boundary conditions adopted near the base of the
convection zone also play an important role: here we have explored
simulations with perfectly conducting bottom boundaries.  Previous
simulations of the solar dynamo \citep[case M3,][]{Brun_et_al_2004}
used a potential field bottom boundary and did not find magnetic
wreaths like those shown here.  When the boundary 
conditions are changed in case~M3 to the perfect conducting
boundaries used here, the axisymmetric magnetic fields grow in
strength, become more organized and form wreaths near the base of the
convection zone.  A variant on that solar dynamo simulation is shown
in the contribution by Miesch in these proceedings.  Conversely, magnetic
wreaths are difficult to achieve in the rapidly rotating dynamos when
we use the boundary conditions of \cite{Brun_et_al_2004}.  
We take some comfort in the fact that wreaths continue to form in
simulations which include a portion of the stable radiative zone
within the domain.  In the simulations with model tachoclines, the
wreaths of magnetism are essentially unmodified from those found in
the simulations which only capture the convection zone.  This suggests
that a perfectly conducting bottom boundary may mimic the presence of
a highly-conductive radiative zone below the convection zone better
than a potential field extrapolation does.

\section{Overview}
\label{sec:challenges}

Stellar convection spans a vast range of spatial and temporal scales
which remain well beyond the grasp of direct numerical simulation, and
we remain humbled by the complexities posed by highly turbulent
convection on global-scales in rotating, stratified, and magnetized plasmas.  
Stellar dynamo studies must drastically simplify the physics of the stellar interior:
as an example, molecular values of $\eta$ in the solar convection
zone range from roughly $10^{2}$--$10^{5}\:\mathrm{cm}^2/\mathrm{s}$ as one moves from
the tachocline to the near surface regions, while the molecular
viscosity $\nu$ is of order $10^{\phantom{5}}\mathrm{cm}^2/\mathrm{s}$ there.
In contrast, our simulations employ values of $\eta$ and $\nu$ that
are of order $10^{12}\:\mathrm{cm}^2/\mathrm{s}$; this 
large value is more similar to simple estimates of turbulent diffusion
associated with granulation at the surface where 
$\nu_\mathrm{t} \sim V_\mathrm{t} L_\mathrm{t} \sim 10^{11}\:\mathrm{cm}^2/\mathrm{s}$ 
given $V_\mathrm{t} \sim 1\:\mathrm{km}/\mathrm{s}$ and $L_\mathrm{t} \sim 1\:\mathrm{Mm}$.
Despite this daunting separation in parameter space, it is striking
that coherent magnetic structures can arise at all in the midst of
turbulent convection.  We find the combination of global-scale spatial
organization and cyclic behavior fascinating, as these appear to be
the first self-consistent 3D convective stellar dynamos to achieve
such behavior in the bulk of the convection zone.

A variety of wreath-building dynamos have been found in these
simulations of rapidly rotating suns:
some build persistent wreaths, while others undergo significant time
variations including quasi-cyclic reversals of global-scale magnetic polarity.
The parameter space is complex, and some simulations show cycles in one
hemisphere but not the other.  Cyclic cases like case~D5 can even
wander into and then back out of non-cycling states.  The role of
tachoclines in stellar dynamos remains a matter of great debate.
These simulations suggest that, at least in rapidly rotating stars,
tachoclines may not play as crucial a role in the organization and
storage of the global-scale magnetic field as in the solar dynamo.
A major step forward will be to explore simulations that couple
wreath-building dynamos in the convection zone, through a tachocline
of shear, to the stable radiative interior below; those simulations are ongoing
now.  We are also exploring how convection and dynamo action may be
different in other solar-type stars, including K-type dwarfs and the
fully convective M-type stars, which present their own mysteries
\citep[see][and contribution by Browning, these proceedings]{Browning_2008}.  
The future of 3D
dynamo simulations is very bright, and cyclic solutions are beginning
to appear in a variety of situations
\citep[e.g.,][]{Ghizaru_et_al_2010, Kapyla_et_al_2010, Mitra_et_al_2010}.
These are exciting times indeed for stellar dynamo theorists!

This research is supported by NASA through Heliophysics Theory
Program grants NNG05G124G and NNX08AI57G, with additional support for
Brown through the NASA GSRP program by award number
NNG05GN08H and NSF Astronomy and Astrophysics postdoctoral fellowship
AST 09-02004. CMSO is supported by NSF grant PHY 08-21899.
Miesch was supported by NASA SR\&T grant NNH09AK14I.
NCAR is sponsored by the National Science Foundation.
Browning is supported by research support at
CITA.  Brun was partly supported by the Programme National Soleil-Terre
of CNRS/INSU (France), and by the STARS2 grant from the
European Research Council. The simulations were carried out with NSF
PACI support of PSC, SDSC, TACC and NCSA.

\newcommand\aap{\emph{A\&A}}%
\newcommand\apj{\emph{ApJ}}%
\newcommand\apjl{\emph{ApJ}}%
\newcommand\jfm{\emph{J. Fluid Mech.}}
\newcommand\an{\emph{Astron. Nachr.}}
\bibliographystyle{apj}
\bibliography{bibliography}

\begin{discussion}

\discuss{T. Rogers}{ 
Could the poleward propagation of field be likened to the dynamo wave,
in which the sign of propagation depended on helicity associated with
convection versus radiation zones?
}

\discuss{B. Brown}{ 
Maybe. Here we clearly see signs that the poleward propagation is
partly due to Maxwell stresses and a resulting poleward slip, and this
phenomena occurs essentially unmodified in simulations that include a
tachocline and convective penetration into a stable radiative
zone. But there may be dynamo wave aspects to the reversals as
well. 
[Note: subsequent work indicates that a non-linear dynamo wave
 does play an important role in the reversal process; see \citet{Brown_et_al_2010b}.]
}

\discuss{E. Zweibel}{ 
Why is the ``failed dynamo'' region on the $\eta$--$\Omega$ plane
localized near the solar rotation rate (\mbox{$\Omega = \Omega_\odot$})?
(see Figure~\ref{fig:parameter space}$a$)
}

\discuss{B. Brown}{ 
On either side of the solar rotation rate, the latitudinal shear is
strong.  Simulations spinning slower than the Sun have strong
anti-solar differential rotation (retrograde equators, prograde poles)
while the more rapidly rotating simulations have stronger solar-like
differential rotation (prograde equators, retrograde poles).  Both
cases lead to higher magnetic Reynolds numbers associated with the
latitudinal shear. Simulations at the solar rotation rate have weaker
differential rotation, and need lower values of $\eta$ to achieve
dynamo action in these wreath-building dynamos.
}


\end{discussion}

\end{document}